# Can Artificial Intelligence Accelerate Technological Progress? Researchers' Perspectives on AI in Manufacturing and Materials Science

John P. Nelson[*,1,2], Olajide Olugbade[2], Philip Shapira[†,2,3,4], & Justin B. Biddle[†,2,5]

***Abstract*—Artificial intelligence (AI) raises expectations of substantial increases in rates of technological progress, but such anticipations are often not connected to detailed ground-level studies of AI use in innovation processes. Accordingly, it remains unclear how and to what extent AI can accelerate innovation. To help to fill this gap, we explore and assess results from 32 interviews with U.S.-based academic manufacturing and materials sciences researchers experienced with AI and machine learning (ML) techniques. We found that AI was primarily used for modeling of materials and manufacturing processes, facilitating cheaper and more rapid search of design spaces for materials and manufacturing processes alike. Benefits included cost, time, and computation savings in technology development. However, AI/ML tools were unreliable outside design spaces for which dense data were already available; they required skilled and judicious application in tandem with older research techniques; and concerns were raised about the potential to detrimentally circumvent opportunities for disruptive theoretical advancement. Based on these results, we suggest there is reason for optimism about acceleration in sustaining innovations through the use of AI/ML; but that support for conventional empirical, computational, and theoretical research is required to maintain the likelihood of further disruptive advances in manufacturing and materials.**

[*]Corresponding author
[†]Equal credit as senior (last) author
[1]School of Public Policy, Oregon State University, Corvallis, USA
[2]Jimmy and Rosalynn Carter School of Public Policy, Georgia Institute of Technology, Atlanta, USA
[3]Manchester Institute of Innovation Research, University of Manchester, Manchester, UK
[4]Alan Turing Institute, Manchester, UK
[5]Ethics, Technology, and Human Interaction Center, Georgia Institute of Technology, Atlanta, USA

Emails: nelsoj34@oregonstate.edu, oolugbade3@gatech.edu, pshapira@gatech.edu, justin.biddle@pubpolicy@gatech.edu
ORCIDs: 0000-0002-3010-2046, 0000-0002-0627-5668, 0000-0003-2488-5985, 0000-0002-8149-5759

## I. Introduction: Artificial Intelligence as an Accelerant to Research and Development?

In recent years, improvements in the power and accessibility of artificial intelligence (AI) have driven increasing attention to AI's applications and implications for scientific research and technological development (Royal Society, 2024). Expectations have been raised that AI will help to accelerate scientific advancement and technological development, leading to economic growth and the development of solutions to global problems. The Organization for Economic Cooperation and Development (OECD) in particular has framed this hope as a way to counteract a perceived slowdown in science, which is taken as a cause of slowing technological development and economic growth (Bianchini et al., 2026; Organization for Economic Cooperation and Development, 2023).

How plausible are such expectations? And how, if at all, could they be realized? A large amount of valuable theoretical and empirical work has already been conducted on the role and effects of AI in science. One strand of literature has focused on proofs of concept or surveys of practitioners on AI applications in science, including for such tasks as phenomenon modeling, literature review, and formulation and testing of hypotheses (Gottweis et al., 2025; Jumper et al., 2021; Mullainathan & Rambachan, 2024; Van Noorden & Perkel, 2023; Xu et al., 2021). Bibliometric research, supplemented by simulation, has provided high-level accounts of the relative affordances of AI in science, tracking output and impact increases; suggesting increases in the importance of computational and of testing capacity; and showing mixed evidence on novelty (Agrawal et al., 2024; Besiroglu et al., 2024; Bianchini et al., 2026; Bianchini et al., 2022; Cockburn et al., 2019). Theoretical arguments both optimistic (Krenn et al., 2022) and less so (Felin & Holweg, 2024; Messeri & Crockett, 2024) have been offered about AI's capacity to offer novel scientific understanding—though scientific understanding is neither the same thing as, nor a simple driver of, technological progress (Bonvillian, 2018; Godin, 2006; Nelson et al., 2011; Sarewitz, 1996 [2010]). A rare ethnographic study has suggested that automation in science creates as well as displaces scientific labor tasks (Ribeiro et al., 2023).

This literature has advanced understanding of how AI is being used in science, and its effects on indicators of scientific quality, scientific resource usage, and labor. Prior literature has also offered various arguments as to the ability of AI to advance scientific understanding. But the core question of whether and how AI can facilitate technological progress has not been answered, and indeed has been little addressed. Answering this question will require investigating not only what AI is being used for, nor only its effects on near-term indicators of scientific quality, but how AI applications and affordances affect fundamental components of the process of technological change. Discussions of AI in science have more assumed than explained how adoption of AI in research and development shifts this process (e.g., Bianchini et al., 2026; Van Noorden & Perkel, 2023). Even conceptual discussions of AI's effects on scientific understanding (e.g., Krenn et al., 2022; Messeri & Crockett, 2024) have not addressed the role of understanding, or AI alteration thereof, in technology development. This gap must be filled if researchers, managers, and policymakers are to make informed, strategic decisions about how to maintain and advance innovation capacity in the age of AI.

Moreover, to date, few studies have undertaken detailed, qualitative investigation of AI's role in innovation processes. High-level survey data (Van Noorden & Perkel, 2023), collections of brief anecdotes (Krenn et al., 2022; Organization for Economic Cooperation and Development, 2023), and bibliometric measures (Bianchini et al., 2026; Bianchini et al., 2022) provide valuable information on impressions, instances, and outcomes, but can shed little light on fundamental mechanisms of technological change. Purely conceptual discussion (Felin & Holweg, 2024; Messeri & Crockett, 2024) or simulation (Agrawal et al., 2024) can frame inquiry and generate hypotheses, but requires empirical validation. Mechanistic, theoretical understanding grounded in detailed empirical observation will be required to draw inferences about when, why, and how phenomena observed through, for example, surveys or bibliometric analyses should and should not be expected to hold and generalize. At present, the paucity of grounded inquiry leaves it unclear exactly how AI could improve rates of technological progress, let alone whether it is managing to do so durably or sustainably. To investigate AI's effects on mechanisms of technological advancement, we report on an in-depth interview study of 32 U.S.-based academic researchers in materials science and manufacturing (MMS) who are using AI or machine learning (ML) in their research.

Unlike most prior inquiries, we frame the question of AI's relationship to technological progress within longstanding evolutionary accounts of technological change. In this framework, science is not a linear driver of technological progress, but an "overhead" field of proxy experimentation and methods development neither sufficient nor, indeed, necessary for improvement (Dosi & Nelson, 2018; Nelson, 2023a; Nelson et al., 2011; Toulmin, 1966). Within this framework, we ask 1) whether AI improves the breadth, reliability, or cost at which new technological alternatives can be tested; and 2) whether it is taking over research tasks previously performed by humans. The answer to both questions is a qualified "yes."

Our results deepen and refine prior suggestions that AI may help to speed up science, but may raise epistemic risks, by clarifying the role of AI-enabled science in processes of technological change; identifying mechanisms and boundary conditions underlying both affordances and downsides of AI; and drawing out implications of these mechanisms for research strategy and policy. Our key, novel findings are as follows: 1) AI's contributions to technological progress come by extending prior computational utilities in reducing the "experiment intensity" of search through technological design spaces; 2) this utility is likely to facilitate sustaining innovation within existing "technological paradigms" (Dosi, 1982; Dosi, 2007; Nelson, 2008b), but may undercut disruptive innovation; and 3) even sustaining innovation facilitated by AI depends upon integration within a robust ecosystem of conventional experimentation, data collection, analysis, and theory development.

Consequently, at least thus far, AI/ML is more another tool in the research and development toolkit than a radical transformation of the innovation process. Conventional experimental and theory development methods remain essential. We suggest that expectations of, and investment in, AI/ML for R&D should, at most, maintain a tempered and contextual optimism. AI/ML investment and use in R&D should be integrated within a robust portfolio of support and funding for more conventional methods and skills development. In addition to their direct implications for the theory and practice of innovation strategy, we hope our results illustrate that reliable, general, and context-sensitive inference about AI's uses and effects in science requires situation of empirical findings within durable theoretical understanding of the intertwined processes of scientific and technological change.

Section 2, below, reviews the burgeoning theoretical and empirical literature on the effects of AI in science, noting that both theory about the connections between AI, science, and mechanisms of technological change, and direct investigations of the effects of AI on technological change, have been scant. Accordingly, this section then reviews theory distilled from the economics, sociology, and history of technological change about how innovation works, and how past scientific advances have facilitated the process of innovation. Section 3 describes our semi-structured interview method. Section 4 reports detailed qualitative results, illustrating that interviewees are largely using AI to extend the preexisting utility of computational modeling—and its attendant upsides and downsides—into additional domains of research. Section 5 situates our findings within the prior literature, arguing that AI's contribution to technological change comes primarily another method of "vicarious trial," and discussing what this theoretical situation suggests about AI's affordances and downsides. Section 6 concludes with limitations, directions for future research, and implications for research strategy and policy.

## II. Background: Artificial intelligence, Research and Development, and Technological Progress

### *2.1 Artificial intelligence and technological progress: What do we know?*

Practical applications of AI (at least, referred to as such) in research and development are relatively novel. There has been a great deal of recent discussion about possible effects of AI for science and for technological development. Understandably, a large amount of attention has been addressed to the question of AI's effects on scientific understanding, particularly from a philosophical perspective. Discussions rooted in philosophy of science have raised hopes have been raised that AI can facilitate scientific functions including data collection, pattern recognition, and hypothesis formulation (e.g., Krenn et al., 2022). Simultaneously, social-epistemological critiques have warned of a degradation or circumvention of human knowledge practices via overreliance on AI (e.g., Felin & Holweg, 2024; Messeri & Crockett, 2024).

As yet, however, empirical evidence is limited, particularly on the specific topic of AI's relationship to technological change. One strand of existing literature focuses on demonstrating the potential of AI systems to augment or take over scientific tasks. Google's report of its AlphaFold protein folding model (Jumper et al., 2021) is a strong example of this genre, as is Google's demonstration of the hypothesis generation capabilities of its LLM-based "AI co-scientist" (Gottweis et al., 2025). However, although media reports confirm substantial investments in AI models for life sciences research, at the time of writing, "no drugs discovered by AI have been approved and many have failed clinical trials" (Heikkilä and Kuchler, 2025). Of course, this situation may change as AI models evolve and more experience is gained in their use. But it does suggest that more advancement will be required if AI is to advance discovery in the life sciences, and perhaps in science in general.

In an analogous strand of work, Bouschery and colleagues (2023) demonstrate uses of LLMs for product development tasks including text summarization, sentiment analysis, and idea generation. Ludwig and Mullainathan (2024), Mullainathan and Rambachan (2024), and Manning and colleagues (2024) all demonstrate prospective approaches for automating generation and/or testing of social science hypotheses using machine learning and large language model techniques. The Organization for Economic Cooperation and Development (2023) and Xu and colleagues (2021) provide illustrative reviews of further such demonstrations across tasks of literature review, hypothesis formulation, data collection and analysis, and hypothesis testing, and Krenn and colleagues (2022) provide anecdotal evidence for such applications from scientists. Yet, such papers provide at most proofs of concept, not strong demonstrations of real-world AI contributions to innovation. An attitudinal survey conducted by the journal *Nature* illustrates widespread adoption of AI among scientists, beliefs in efficiency gains, and concerns about misuse, but does not explore why, when, or how each of these tendencies should be expected to manifest or persist (Van Noorden & Perkel, 2023).

For a more general view, bibliometric studies that seek to connect the use of AI with scientific progress and innovation are emerging. For example, Cockburn and colleagues (2019) demonstrate heterogeneity in adoption of AI across scientific disciplines, but they do not attempt to show whether AI adoption improves research. In a bibliometric study of neural network adoption in health science articles, Bianchini and colleagues (2022) find neural network adoption to associate negatively with recombinant novelty and positively, though with high spread, with article impact. In a more recent study, Bianchini and colleagues (2026) find AI adoption in general to associate positively with scientific article impact and novelty, though with high heterogeneity across scientific fields. These results are modest and not altogether promising regarding AI's potential to accelerate even scientific progress. Agrawal and colleagues (2024) use a pure simulation model of hypothesis generation and testing to suggest that scientific gains from AI will depend heavily upon testing capacity. Based on analysis of large datasets of articles on machine vision, Besiroglu and colleagues (2024) find AI to increase the research productivity of computational capital, which they attribute to advantages in idea production.

In a rare qualitative study of automation in science, Ribeiro and colleagues (2023) draw on ethnography, surveys, and semi-structured interviews to find that automation and digitalization in biosciences labs generates large amounts of novel "mundane knowledge work" in data wrangling and automation servicing and maintenance. These results complicate the assumption that AI and other forms of automation are necessarily labor-saving or that they reduce the incidence of mundane or routine tasks.

Concerningly, the literature on effects of AI in science largely takes for granted a straightforward connection between scientific production of "ideas," technological progress, and economic growth (following the lead of Bloom et al., 2020; and indeed of Bush, 1945), whereas these interactions are actually highly complicated and contingent (Block & Keller, 2011; Bonvillian, 2018; Dosi & Nelson, 2018; Godin, 2006; Nelson, 2023a; Toulmin, 1966). In brief, more science does not always improve understanding, nor does even better science by bibliometric measures (Forscher, 1964; Hicks et al., 2015; Ravetz, 1971; Sarewitz, 1996 [2010], 2016). Even novel scientific understanding does not always lead to technological progress (Bozeman & Sarewitz, 2011; Nelson, 2026; Nelson & Bozeman, 2025; Sarewitz, 1996 [2010]; Sarewitz et al., 2012; Sarewitz & Pielke, 2007). And sometimes technological progress advances independently of, or drives, novel scientific understanding (Hughes, 1989; Misa, 1985; Nelson, 2023a; Nelson et al., 2011; Rosenberg, 1974; Stokes, 1997; Toulmin, 1966).[1] Determining when, where, and how scientific productivity leads to technological advancement, and when and how AI can facilitate such, requires engagement with the extensive prior literature on processes of technological change.

*2.2 How does technological progress work?*

Decades of research in the history, sociology, and economics of innovation show that development of improved technology is a process of iteratively developing, adapting, spreading, and replacing both physical artifacts and routines of human practice and organization (Nelson et al., 2011; Vincenti, 1990). This iterative process is often described as evolutionary, with all the complexity, dynamism, and contingency this term entails (Dosi & Nelson, 2018; see generally Gould, 2002). A brief elaboration of this account will help to situate AI's potential utility. Human societies improve our "know-how" by developing new ways of doing things (including modes of thinking, such as both implicit and explicit theories), trying them out, and judging whether these new practices better serve their goals than did old practices. New technological variations are developed; they survive at different rates; surviving technologies propagate and go on to vary in turn. This overall process is as old as human tool use, but the modes of iteration, trial, and evaluation available to humanity have expanded enormously in the last few centuries (Hughes, 1989; Mokyr, 1990, 2002; Rosenberg & Birdzell, 1986). Mechanisms of variation, differential survival, and propagation vary from the intentional and explicit (design, deliberate trial, purchase of new technological artifacts) to the systemic and emergent (drift in practice, systemic selection, informal interchange within "invisible colleges" [Crane 1972] of practitioners) (Figure 1).

For an example of change in the process of technological evolution, Vincenti (1990) tracks how the development of aeronautical theory and computational technologies over the 20th century made it possible for aeronautical engineers to substitute first scale models in wind tunnels, then entirely computational models, for preliminary trials of new airfoil designs. These methods of indirect, "vicarious" trial cost less time, resources, and danger to pilots than did test-flying every possible design; and permitted more comprehensive observation of new airfoils' properties. Similarly, once, there was no trial available for prospective medicines but putting them in a human body and seeing what happened. Now, researchers have animal trials, petri dishes, and even computational analyses of the chemical dynamics of proteins (Yaqub, 2018; Yaqub & Nightingale, 2012).

These preliminary trials did not eliminate more intensive modes of trial. Rather, they facilitated early-stage search for promising alternatives to advance to more intensive trial modes. No rocket or airplane is treated as reliable until it has flown and landed safely, many times and under many conditions. No drug is treated as safe and efficacious until it has passed human clinical trials in a diverse patient body. But, in modern, industrial knowledge ecosystems, there are many useful preliminary trials available *en route* to these final validations.

*2.3 How does science contribute to technological progress?*

The role of science in technological change can be further clarified through a complementary theoretical idiom also drawn from the innovation studies literature. What can be described as a process of "evolution" from a historical perspective is a process of search for superior alternatives on the level of individual technology developers. In recent decades, this understanding has been refined into a view of science as a "map" (or, we would suggest, a set of partially overlapping maps, of various degrees of granularity, breadth, and reliability) in the process of search for novel and useful technological alternatives (Agrawal et al., 2024; Fleming & Sorenson, 2004; Nelson, 2008a; Nelson & Winter, 1982; Sorenson, 2024). That is, science does not replace the longstanding process of practical trial and error in efforts to develop useful and novel practices and artifacts; but it supplements and guides this process.

[1] Though it is not a major focus of this paper, it is worth mentioning that the final step from technological advancement to economic growth, and the implicit step from economic growth to societal prosperity, is no simpler. Innovations in extractive, exploitative, destructive, tyrannical, resilience-reducing, or merely distracting technologies can, rather than advance, reshape, displace, or even undercut both economic prosperity and human well-being (Bozeman 2020; Mazzucato 2018, 2021; McCraw & Childs, 2017; Nelson & Bozeman, 2025; Scott, 1998).

**Figure 1:** A schematic depiction of the process of technological evolution. Figure adapted from Nelson (2023b).

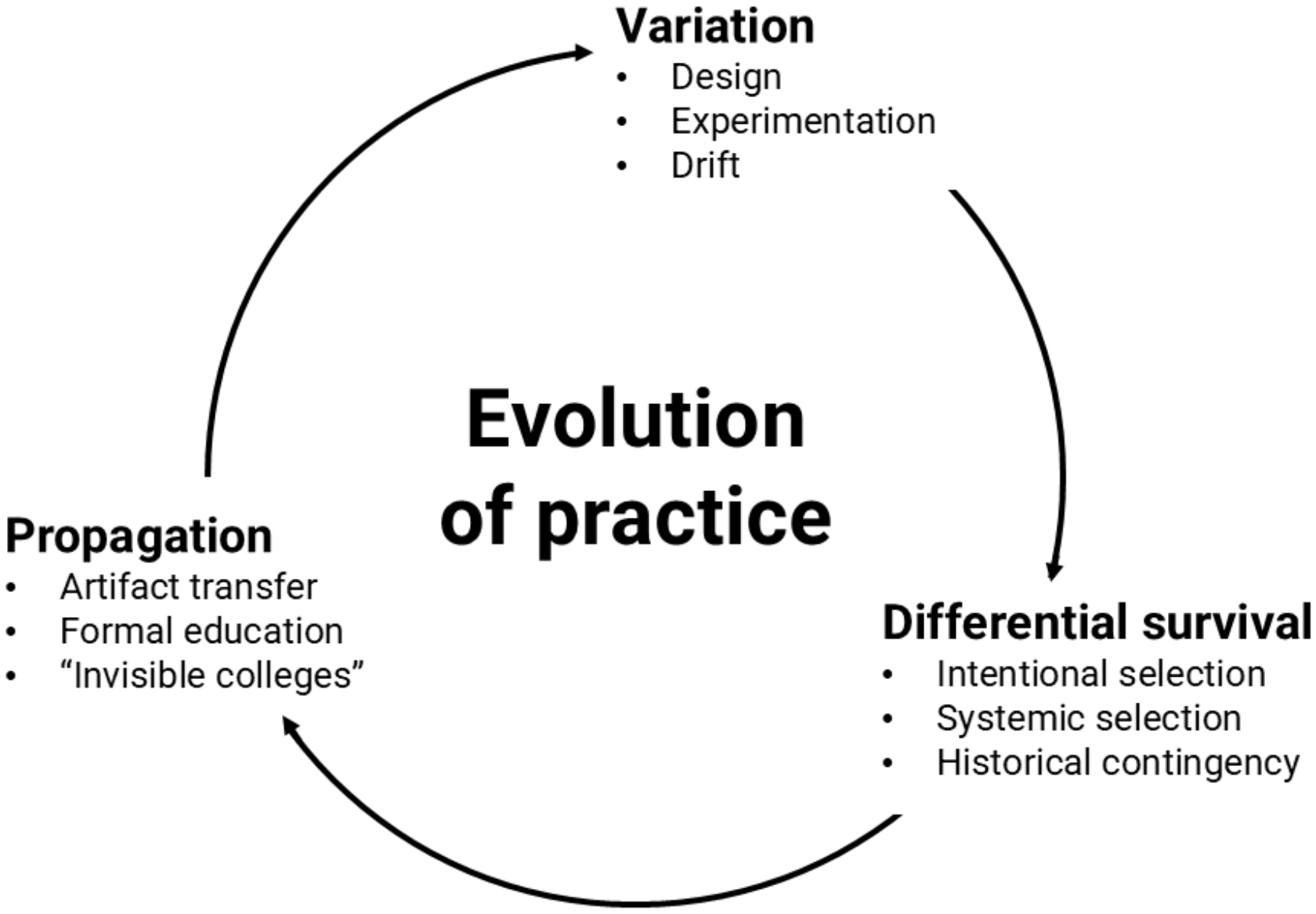

Science suggests to technology developers potential artifacts or practices which may prove useful, and, equally helpfully, suggests possible avenues of investigation which may not (Hughes, 1989). This is "vicarious trial"; it is cheaper and easier to explore the landscape, and choose one's preferred location, by looking at a map than by walking the region's full extent. Informal, tacit, experiential knowledge has always served this function, but the suggestions offered by formal, explicit scientific inquiry and theory can often reach further from prior experience; and be more easily transmitted (Collins, 2010; Fleming & Sorenson, 2004; Hutchins, 1995). In so doing, science can suggest and support pursuit of more novel capabilities, save effort and resources, and provide reason to pursue possibilities which may seem at the time very dubious. In this sense, science serves as a sort of intellectual (as well as physical) tool-production industry for technology developers (and to various practitioners of other sorts). Scientific tools become increasingly necessary for innovation as technological systems become more elaborate, infrastructurally embedded, and tightly coupled (Fleming & Sorenson, 2004).

The "maps" provided by science have grown steadily wider, more reliable, more detailed, and more parsimonious in many areas of human enterprise in recent centuries, and many formerly separate maps have been at least partially stitched together (Nelson, 2023a; Vincenti, 1990). Maps of mapmaking, and maps of those maps, have been developed, in the forms of the various branches of mathematics and statistics, philosophy of science, and science and innovation studies (Ravetz, 1971; Saltelli et al., 2020). Yet the maps constituted by scientific (and other sorts of formal) knowledge are eternally incomplete, fragmented, and ragged--at the edges, in the interstices, and at the apparently inexhaustible frontiers of both decreasing and increasing scale (Cartwright, 1999; Dupré, 1995).

Philosophical and social epistemologists of diverse persuasions agree that science consists at least in large part in induction and extrapolation from prior observation and experience into formal, explicit concepts and theories (Allen, 2004; Collins, 2010; Dewey, 1916 [2004]; Dupré, 1995; Rorty, 2021). In many areas, such extrapolation can reach far indeed; but it is always rooted in what has previously been experienced and observed (and recorded, and reflected upon, and analyzed, and systematized; Allen, 2004; Laudan, 1977; Popper, 1934; Simon, 1983). New technologies are made by putting older technologies (and, sometimes, previously unexploited materials) together in new ways, and the same might be said of new ideas. Recombinant search is always local, though science can expand the size of the available locality (Allen, 2004; Kline, 1995; Kuhn, 1962 [1996]; Rorty, 2021). Science's dependency on observation explains the strong association between novel observational or experimental apparatus; and disruptive advance in science. New observational tools provide new sorts of observations to work into the maps (Krauss, 2026).

Development and extension of increasingly reliable "cartographic practices," or methods of vicarious trial, has been characteristic of technological fields which have seen rapid progress in recent centuries – fields such as metallurgy, manufacturing, weapons systems design, drug-based treatment of infectious disease, or telecommunications and information processing. These more rapid and often less costly methods of development and trial speed the process of search for superior designs and routines, thus, accelerating the rate of technological advancement (Nelson, 2023a, 2023b; Nelson, 2011). If AI is to accelerate technological progress, this is the process it will have to improve. This analysis aligns with Agrawal and colleagues' (2024) suggestion that gains from AI in research will depend substantially upon testing capacity.

### *2.4 How could artificial intelligence increase the rate of technological progress?*

In a review of literature on mechanisms affecting rates of technological progress, Nelson (2023a) identifies seven characteristics of technological fields which make them easier or harder to advance. Of these, AI appears directly relevant to two. The first, already discussed extensively, is the availability, reliability, and cost of Vincenti's (1990) "vicarious trial." The cheaper, more available, and more widely applicable methods of vicarious trial, the cheaper and faster is the process of trying out new technological alternatives and sorting the better from the worse. As already noted, prior literature demonstrates that AI systems may, in some domains, be able to generate or test viable technological alternatives more rapidly or less expensively than could previous workflows (Gottweis et al., 2025; Jumper et al., 2021; Manning et al., 2024).

The second AI-relevant variable affecting the difficulty of technological progress is what Nelson calls "artifactuality," i.e., the extent to which core task processes are performed by nonhuman artifacts. This argument follows on the longstanding observation in sociology and history of technology that nonhuman artifacts can do things which human beings cannot, can be altered to better fit task processes in ways human minds and bodies cannot, can be more easily compared to one another than can human task processes, and can often be reproduced and disseminated more reliably than can human skills (Marx, 1936 [1867]; Mokyr, 1990; Nelson & Nelson, 2002; Rosenberg, 1974). This argument applies to AI in two ways. First, if AI displaces human practices in some fields of technology (e.g., quality control of microchips), such AI systems may be more readily, reliably, and ethically modified and improved for their roles than can human beings. However, this is not a result of use of AI in R&D – rather, it is reason to expect that R&D on AI may be easier than R&D on human workers.

Focusing on use of AI in R&D itself, to facilitate progress through artifactuality, AI would need to take over research tasks themselves. In so doing, it might be able to suggest different and/or more reliable technological alternatives than can human practitioners; execute trial processes in ways superior to humans; or disseminate improved technologies in ways superior to humans. All of this falls under the overall heading of facilitating faster, cheaper, or more generally applicable vicarious trial. Accordingly, in the context of R&D, the artifactuality question is a specific subset of the vicarious trial question.

The reviewed theory on mechanisms affecting the rate of technological progress informs two primary research questions, one for each identified mechanism.

> **RQ1:** Is AI improving the breadth, reliability, or cost of vicarious trial?
>
> **RQ2:** Is AI taking over research tasks previously performed by humans?

Due to the paucity of grounded knowledge about AI's deployment and effects in research and development, we investigated these questions via a semi-structured interview study with practicing researchers in manufacturing and materials science (MMS). Because this research is early-stage, exploratory, and qualitative, we do not offer hypotheses.

## III. METHODS

### *3.1 Case: Artificial intelligence in manufacturing and materials science research*

Manufacturing and materials science are well-developed, practically oriented research domains. They serve as a bridge between multiple long-established industries, still foundational to all industrial society, and some of the most powerful and well-established sciences (in particular, physics and chemistry). MMS researchers in academia are well-supported by both private and public funding, and they frequently collaborate closely with private firms on practical problems of design and production. All of our interviewees explicitly stated that their research, even when theoretical, was intended to improve specific technologies, often in partnership with firms or government (including defense) organizations. Moreover, many of the advanced mathematical, statistical, and computational tools which have been rolled under the heading of AI in the last decade have long histories in MMS, making MMS researchers well-situated to benefit from novel AI capabilities. In short, MMS provides an opportunity to investigate the effects of AI in R&D in a well-established, practically oriented field drawing on longstanding and reliable scientific disciplines and theories.

As discussed further below, our sampling method aimed for manufacturing researchers, and this frame included a wide array of formal disciplinary specializations, including mechanical engineering, materials science, industrial engineering, computer science, and more (Appendix 2). The line between manufacturing process design and materials design has been increasingly blurred over time, with a substantial body of contemporary research aiming to optimize both simultaneously. This is in part because, in cutting-edge manufacturing methods such as additive manufacturing, the production process has great consequences for the microstructure and performance of the material. For example, in additive manufacturing of aerospace parts, the grain size of different metals in an alloy and the porosity of the part are substantially determined by such details as whether the part is produced via laser powder bed fusion or by blown powder directed energy deposition, the grain size of the powder, the flow rate of the powder, the speed of movement of the laser or deposition head, and the intensity of the laser. Meanwhile, alloys are, in part, selected because they are expected to deposit well and yield desirable performance characteristics under particular ranges of these process variables. The material, the process, and indeed the part geometry cannot be separated; all are often treated in one and the same investigation.

Our sampling method used manufacturing-related keywords and conferences identified by an informant as loci of manufacturing research. We refer to the overall field of addressed research as "manufacturing and materials science" (MMS) to communicate that a large subset of the researchers interviewed considered themselves to be materials scientists focused on problems of manufacturing.

*3.2 Method: Semi-structured interviews with practicing academic researchers using artificial intelligence*

Because of the limited prior empirical and theoretical literature on AI in science, we chose a research method appropriate for elucidating novel phenomena and constructs of interest in little-understood areas of inquiry: semi-structured interviews. Semi-structured interviews prescribe topics to address but also leave space for interviewees to go off-script to follow up on important points. Our semi-structured interview protocol (Appendix 1) was designed based upon constructs which have already received attention in the literature. It aimed to investigate a variety of topics in AI research, including but not limited to those discussed in this paper. As AI is a fuzzy term, and as not all relevant researchers consider themselves to work with it, we framed our interview protocol to focus on AI and machine learning (ML). This protocol focused on the following topics: researchers' personal experiences with AI/ML in research; reasons for using AI/ML in research; advantages and disadvantages to AI/ML in research; effects of AI/ML on resource and skill needs of research; and effects of AI/ML on research communications.

In this paper, we address only results relevant to the research questions articulated in the Background section above. For our sample, we aimed to engage researchers in the field of manufacturing and materials science with experience using AI/ML for research. We chose to focus on academic researchers because such researchers tend to be more open to interviewing about their work, and because they are an essential part of cutting-edge manufacturing research. Due to resource and language limitations, we could not sample researchers worldwide. We chose to focus on U.S.-based universities 1) for reasons of personal access; and 2) because the U.S. is a world leader in artificial intelligence research, particularly in production of English-language articles.

As there is no preexisting database of such researchers, we constructed our own sample frame. We began by identifying relevant researchers through the academic literature database Web of Science (WoS). We identified articles and conference papers on manufacturing and involving AI/ML using a narrowly targeted search crossing several core AI/ML terms with the term "manufacturing," conducted on May 8, 2024:

> TS=("Artificial Intelligen*" or "Neural Net*" or "Machine* Learning" or "Expert System$" or "Natural Language Processing" or "Deep Learning" or "Reinforcement Learning" or "Learning Algorithm$" or "*Supervised Learning" or "Intelligent Agent*") AND TS=(manufacturing)

Due to resource and web documentation limitations, we limited our initial sample frame to researchers at the ten U.S. universities most productive of articles and papers returned for this query from 2019-2024. These universities appear in Table 1.

For each target university, we constructed a list of affiliated authors from the Web of Science. We manually verified their affiliations and collected contact information through their university web profiles, yielding 107 researchers. We also searched each university's websites for research centers, research groups, or topic areas relevant to AI/ML or manufacturing. We manually reviewed faculty and staff affiliated with such centers, groups, or topic areas for additional researchers at the intersection of AI/ML and manufacturing, yielding 89 more prospective interviewees. We solicited interviews from interviewees identified online via three emails, sent one

**Table 1:** U.S. universities most productive of AI/ML manufacturing articles and conference, 2019 – 2024*.

| University | Relevant papers |
|---|---|
| Georgia Institute of Technology | 96 |
| Massachusetts Institute of Technology | 56 |
| University of Michigan | 55 |
| Purdue University | 43 |
| Rutgers University New Brunswick | 37 |
| University of California, Berkeley | 37 |
| Northwestern University | 35 |
| Texas A&M University | 34 |
| Virginia Polytechnic Institute & State University | 30 |
| North Carolina State University | 28 |

*Source: Web of Science, see main text for search approach. *January 1, 2019-May 8, 2024.*

week apart. The third was sent by a colleague who is a prominent materials scientist, in hopes of increasing response rate.

Finally, to ensure our sample was connected to current events in the field, the lead author attended two manufacturing and materials science research conferences, recommended by a manufacturing researcher informant, to identify additional prospective interviewees: the 2024 Materials Science and Technology Conference in Pittsburgh, PA; and the 2024 International Conference on Advanced Manufacturing in Atlanta, GA. This author attended sessions on AI/ML at these conferences and solicited interviews from speakers face-to-face following their presentations, following up after via email. Eighteen interviews were solicited in this way. From these sampling methods, we solicited 214 interviews and completed 34. However, we excluded two interviews from analysis after completing them. This is because, while these two researchers had been authors on AI/ML manufacturing papers, their main research areas were too distant from the topic: smart cities and engineering education, respectively. This exclusion left us with a sample of 32 interviews.

Interviews were conducted via the video conferencing platform Zoom or in person. All interviews lasted between 30 and 75 minutes, save for one which had to be cut short at 15. All interviews were recorded. Interview recordings were automatically transcribed, initially using the web transcription service Otter.ai, later locally on a research team computer using OpenAI's local Whisper tool (this switch was made because Whisper produced higher-quality initial transcripts). Authors listened through each recording and manually corrected errors in the automatically generated transcripts.

We analyzed interview transcripts using an iterative, qualitative coding method, using the qualitative data analysis software NVivo 14. Analysis used a hierarchical coding scheme starting with deductive (theoretically derived, prescribed) top categories and inductive (derived from the data) subcategories; and iterating through the coding process (MacQueen et al. 1998; Miles & Huberman 1994). Analysis "triangulated" between interviewees' comments, i.e., noted points of agreement and disagreement and placed greater confidence in repeated and uncontested views (Tracy, 2010). However, contested or rare themes are still reported, as these can indicate important controversies, points of uncertainty, or underrecognized issues meriting further investigation. To check validity, once we had prepared draft manuscripts, we circulated them to our interviewees with a request that they correct any errors. Their comments were integrated into subsequent drafts.

## IV. Results

### *4.1 Respondent Information*

Table 2 provides response rates for each interviewee identification method. Our initial response of 34 interviewees provides a response rate of 16%; excluding the two interviewees whose research topics were too distant from the focal area drops this to 15%. For a sample solicited primarily via cold emails, particularly of busy professionals such as academic researchers, this is an acceptable response rate. Statistics on cold-call responses to interview requests are not available, but Wu and colleagues report an average response rate of 44.1% across a sample of 1071 online surveys on education yielding peer-reviewed publications. However, respondents to such surveys were not typically academic researchers or other "elite," highly busy respondents; and 20% received compensation.

**Table 2:** Interview solicitations and completions by interviewee identification method.

| Source | Solicited | Completed | Response Rate |
|---|---|---|---|
| Web of Science | 107 | 15 | 14% |
| University Websites | 89 | 12 | 13% |
| Conferences | 18 | 7 | 39% |
| ***Total*** | ***214*** | ***34**** | ***16%**** |

** Two interviews excluded after completion because main research areas (smart cities and engineering education) were too distant from AI/ML in manufacturing.*

For a better comparison, Bozeman and colleagues (2023) report a response rate of 30.3% on an uncompensated survey of academic researchers publishing in major public administration and public policy journals. This is something of a best-case scenario for small-team solicitations for uncompensated research participation by busy professionals, for two reasons. First, a survey is a less laborious request than an interview. In addition, two of the soliciting authors, Bozeman and Bretschneider, were very prominent researchers in the field, boasting nearly 50,000 combined citations and personally known to a large proportion of the target sample.

Meanwhile, a total sample of 32 participants is reasonable for in-depth interviews, which tend to focus on internal detail over breadth. Thirty hours of interviews provide extensive detail on researchers' specific experiences and practices. While literature on interview sample sizes specifically in the context of science studies is scant, this sample size is in line with empirical findings and recommendations for reaching content saturation from health, market, and educational research (Ago & Bekele, 2022; Morse, 2015; Squire et al., 2024; van Rijnsoever, 2017; Vasileiou et al., 2018), as well as with well-regarded interview studies in science and innovation (e.g., Pinch, 1981; Ribeiro et al., 2023).

Also worthy of attention are the characteristics of interviewees (Appendix 2). As the authors were all Georgia Tech researchers at the time of data collection, it is not surprising that the response rate from Georgia Tech personnel was substantially higher than for other institutions. Exploratory analysis revealed no systematic difference between the responses of Georgia Tech-affiliated interviewees and other interviewees. Our sample skewed senior, with full professors making up a plurality of respondents; and mostly male, with women making up only 13% of respondents. For 2023, the American Society for Engineering Education reports that women comprise about 20% of all faculty in U.S. engineering, computing and engineering technology programs and just 15% of full professors (American Society for Engineering Education, 2023). Female representation among our respondents is slightly lower compared to these national statistics although not by much: our focal field, as we also observed from fieldwork in manufacturing labs and conferences, remains predominately male.

Fortunately, the purpose to which we put this sample – exploratory, qualitative research intended to elucidate phenomena and constructs for further inquiry – makes no use of inferential statistics. Our respondents were experienced, mostly high-ranking researchers affiliated with major, accredited research universities and with substantial personal experience using AI/ML in research, lending credence to their ability to elucidate the dynamics of AI/ML in research. Further investigation of the topics discussed here will benefit from larger and more diverse samples, as well as different methods entirely (e.g., surveys, case studies, ethnographies).

### *4.2 Applications of artificial intelligence and machine learning among interviewees*

Materials design, product design, and manufacturing process design have always been intertwined in practice, and they have become increasingly intertwined in research as well. ML tools offer opportunities for integrated, multifactorial analysis of how materials properties, part geometry, and manufacturing processes interact to produce final part properties and production performance via modeling. Many of our interviewees explicitly commented upon the interdisciplinary nature of their work; those who identified as disciplinary mechanical engineers, materials scientists, industrial engineers, or chemical engineers also considered themselves to be manufacturing researchers. Accordingly, we characterize interviewees' subject areas by the aspect of manufacturing on which they primarily focus: manufacturing processes; materials; or final products. Nevertheless, some interviewees fall in the overlaps, and address of any of these problems in contemporary manufacturing requires information from the others (Fig. 2). Exploratory analysis (Appendix 3) revealed no strong heterogeneity in findings across subfields, so the remainder of this discussion treats these subfields together.

Our results (Tables 3 & 4) require some context about the role of computer models in technology development. Broadly speaking, computer modeling permits vicarious trial, as discussed above. Rather than fabricating and evaluating novel compounds or parts or testing new process configurations in the real world, models can permit technology developers to search through design spaces using computation and to select promising options for real-world trial. Theoretically informed mathematical and computational models have served this purpose for decades. However, not all materials, phenomena, or manufacturing processes have accurate, existing theory or computer models. In addition, some models are very computationally intensive to run, limiting their application.

Per interviewees, AI/ML tools offer two broad categories of novel modeling opportunities. First, AI/ML tools can be used as "surrogate models" for preexisting physics-based models. That is, an AI/ML model can be trained on inputs and outputs for a computationally intensive physics-based model, and – once trained – can be used as a sort of "lookup table" to yield similar results to the original model at lower computational cost:

**Figure 2:** Interviewees' areas of focus within manufacturing research.

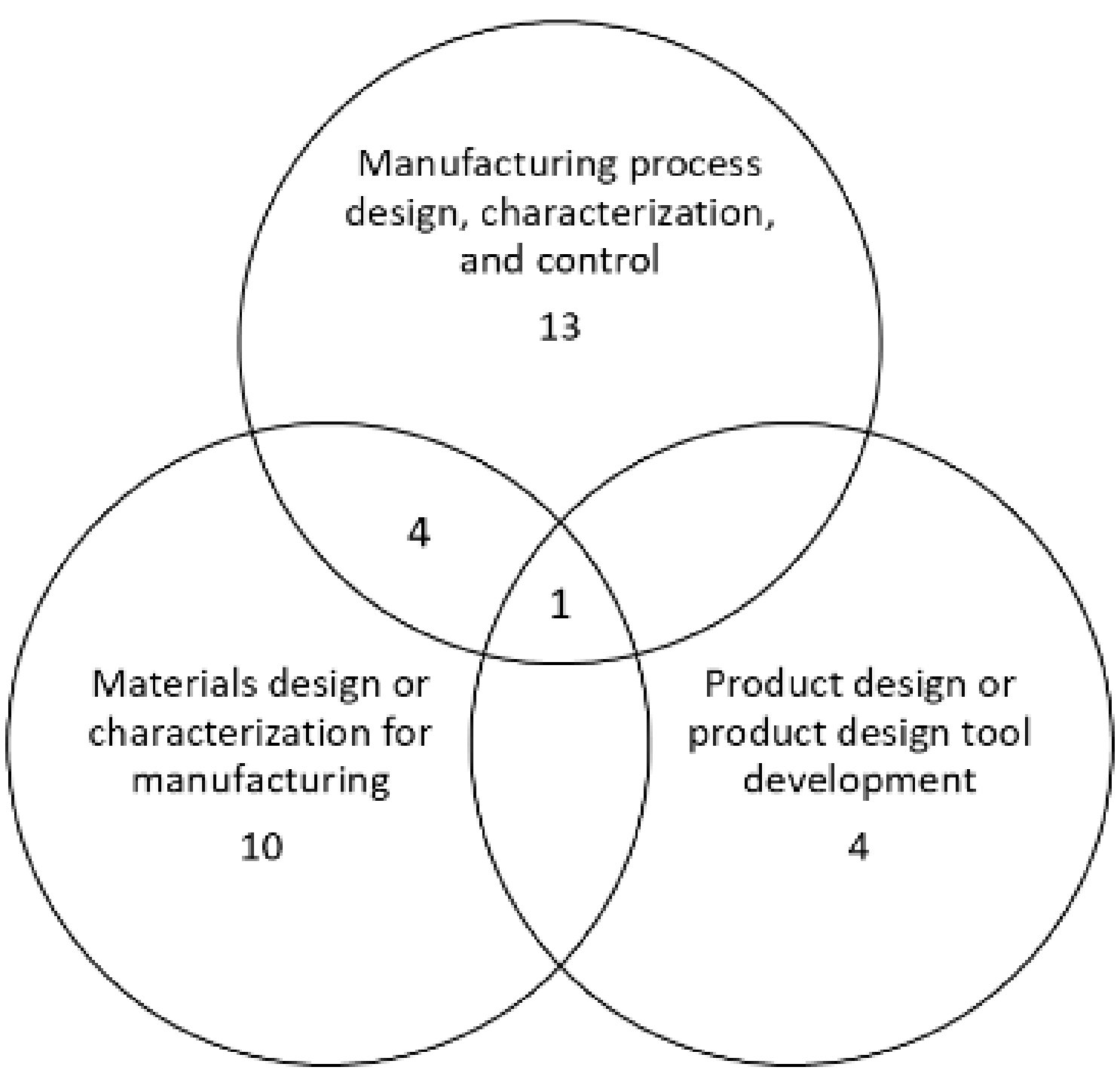

> [W]e use these reduced order models, which are basically AI based representations of the more accurate models that we use . . . It's sort of just doing input-output, rather than doing the actual processing . . . a multi step equation solving or processing all the time. That's what makes it faster (AI10).

The computational savings can be substantial:

> For the computational wise, suddenly you see a lot of speed improvement . . . speeding . . . the prediction by 100 times or even more, and many times could be even more, so . . . ten to the fifth (AI02).

Second, AI/ML models can be trained on input and output data for phenomena (e.g., complex production processes) which lack robust theoretical models, developing novel predictive capabilities in the absence of explicit, human-designed theory. This is sometimes referred to as "phenomenological modeling," as it attempts to model phenomena in the absence of mechanistic, explanatory understanding:

> [T]he first reason we choose to use AI is because we don't have a good model of what our system is. . . I get a bunch of data coming in and I have a bunch of sensor readings, you know. . . And I use the AI to map the bunch of sensor readings to the process health or process status or machine status that I have. And it's very black box and I don't particularly like doing that, but it's quite useful (AI27).

Such AI/ML tools can either be used to address entire phenomena, or to model only the uncertainty remaining within an existing physics-based model. This latter "hybrid" approach uses AI/ML to try to correct for unknowns and uncertainties within an existing model:

> You can also use AI to let the physical model talk to the measurement, experimental data . . . to take into account the unknown parameters . . . The unknown physics not captured by the original physics model (AI29).

These two basic utilities lead to a variety of opportunities and consequences, discussed further below.

Twelve interviewees discussed using AI/ML for design of production processes, e.g., by predicting the effects of specific process parameters on part quality; and selecting parameters based upon these predictions. Four focused specifically on ML-based prediction of part properties based on microstructure. Three interviewees discussed using AI/ML to predict other attributes of part designs, including manufacturability and performance characteristics (such as turbulence around parts). One interviewee discussed a research project aiming to produce a generative AI-aided design tool which would be accessible for consumers. Finally, one discussed using ML-based tools to organize preexisting design information.

**Table 3:** Summary of AI/ML applications among interviewees. All codes nonexclusive (n=32). Granular subcodes depend upon specific, detailed statements, and not all more general statements may also include more specific details.

***Modes of use***

AI/ML modeling of phenomena (24)

- Phenomenological modeling of poorly understood phenomena without explicit theory* (20)
- Surrogate modeling for physics-based models (12)
- Providing correction terms for physics-based models (3)
- Planning experimental campaigns (4)

Measurement of constructs (2)

AI/ML tool as research product (1)

- Generative AI design tool for consumers (1)

Self-driving lab (1)

Large language models & generative AI (15)**

- General literature search and review (11)
- Drafting papers or preparing graphics (7)
- Extracting data from documents (4)
- Assistance with programming (2)
- Translating papers (1)

**A large share of interviewees' statements about modeling which did not explicitly pertain to surrogate modeling likely refer to phenomenological modeling. However, to avoid overinterpretation, we have assigned this code only where the interviewee explicitly discussed using AI/ML to model unknown relationships.*
***The interview protocol did not focus on LLMs, so use may be underreported.*

**Table 4:** Benefits and downsides of AI/ML in research discussed by interviewees. All codes nonexclusive (n=32). Granular subcodes depend upon specific, detailed statements, and not all more general statements may also include more specific details.

***Benefits of artificial intelligence or machine learning in research***

Reduction in accuracy/cost/speed trade-off in research, especially computer modeling (26)

- Reduced computation time (17)
- Replacing experimentation (11)
- Reducing need for computationally intensive, physics-based models (2)
- Saving research labor (6)
- Exploring larger design spaces (3)
- Accuracy improvements (4)

Address of previously unsolvable problems (16)

- Identify human-unidentifiable patterns or phenomena (9)
    - Feature recognition
    - Construct measurement
    - Modeling poorly-understood relationships between variables
- New scales or speeds feasible (4)

***Downsides of artificial intelligence or machine learning in research***

Accuracy weaknesses (24)

- Predict poorly outside regions of dense, high-quality training data (10)

Interpretability weaknesses (11)

- Bounds of accuracy can be unclear (3)
- Accuracy assessment can be difficult (3)

Long-run scientific progress concerns (5)

- AI/ML cannot develop novel scientific theory (5)
- AI/ML may bypass opportunities to identify empirical or theoretical novelties (1)

Resource issues (7)

- Data acquisition and cleaning is time-intensive (3)
- AI/ML models are computation- and energy-intensive to develop (2)

Inappropriate use issues (13)

- Easy to over-trust (7)
- May be inappropriately used to address problems soluble with simpler methods (4)

Eleven interviewees discussed using AI and ML tools for process monitoring and control applications. Seven discussed applications in machine health monitoring, using AI/ML systems to process data collected from machines in operation and draw current inferences or future predictions about machine wear and breakage. Meanwhile, five discussed using AI/ML systems to predict the performance of manufacturing processes based on machine data, and to adjust production parameters to maintain quality or resolve process faults.

Other interviewees discussed using AI/ML as part of the experimental workflow, in particular, among the fifteen materials science-focused interviewees. Several interviewees integrated ML-based tools into the sample characterization workflow, e.g., for translating imaging or elemental analysis data into materials or part properties. On a more general level, several interviewees discussed using AI/ML to plan "campaigns" of experimentation or of conventional physics-based simulation to characterize new design spaces:

> Instead of the traditional, like, let's think about what we want to know . . . and then create a design of experiments, and then execute all the experiments and then make the model. . . . You actually look at the whole space of the problem you're trying to solve, and then pick a place to start with the experiments, and now build the model and update the model in real time as you keep doing . . . batches of experiments. So it's really machine learning-guided walks through a design of experiments landscape. . . (AI09)

> You can sort of think of machine learning as telling you where you need to go, or as a guide. And then, if you really cared about the answer, you would check with a more, you know, high-fidelity approach, like [density functional theory, a computationally intensive, highly accurate physics-based modeling approach], or an experiment. . . So if, for instance, if we're doing, like, a materials discovery exercise, we may screen tens of thousands of materials, with machine learning, and in the end, maybe you do like one hundred [density functional theory] calculations or something on the most promising ones (AI12).

As discussed above, like preexisting physics-based modeling approaches, AL/ML-based modeling approaches can directly substitute for experimentation in some cases. One interviewee discussed using AI/ML systems to run automated experiments, similar to the well-documented phenomenon of self-driving labs (Abolhasani & Kumacheva, 2023; Tom et al., 2024).

Finally, fifteen researchers discussed using AI/ML for applications outside the distinctive tasks of MMS research. Ten discussed using LLM-based tools for general literature search and review, and seven discussed using LLMs or generative image models for help drafting or editing, or preparing graphics, for papers or presentations. Four discussed using LLMs to extract information from large text datasets (in this case, machine manuals and engineering literature). Two mentioned using LLMs for assistance in programming. One mentioned using LLMs to translate papers into English.

*4.3 Benefits of artificial intelligence or machine learning use*

As discussed above, most applications of AI/ML among interviewees involved computer modeling of parts, products, or production processes. Correspondingly, by far the most commonly discussed benefit from AI/ML use, raised by 26 interviewees, was a loosening of the accuracy-speed tradeoff characteristic to computer modeling. AI/ML techniques can often produce similar results to conventional, physics-based simulations at lower computational cost. This reduction in trade-offs can manifest as several different practical benefits. First and most straightforwardly, discussed by 17 interviewees, AI/ML tools can simply reduce the computation time required to achieve a given result, saving on electricity, hardware use, and, often, time. Smaller numbers of interviewees discussed analogous benefits in replacing more experimentation with cheaper and faster computation (8), reducing the frequency with which compute-intensive physics-based models need to be used (2) or generally saving labor involved in research (3).

In addition, several researchers discussed time savings emerging from modern computer science infrastructure, including code libraries and LLM-based coding assistance. A complement to simply reducing the resource intensity of some modes of modeling is by using more efficient computation to explore larger design spaces than would be feasible using conventional, physics-based simulation, discussed by three interviewees. However, not all interviewees were so sanguine. Nine stated that, while AI/ML-driven resource savings for research may materialize in the future, they have not done so yet in their own research.

Another reported benefit, highlighted by sixteen interviewees, is that AI/ML tools permit solution of research or engineering problems which could not be solved before. AI/ML can contribute to such novelties in several different ways. Perhaps simplest is efficiency improvements making viable some research projects or programs which would not have previously been feasible:

> [L]ike a lot of chemistry up until now, things have remained the same for so long, because, at the end of the day, it is more cost-efficient or even mainly just time-efficient to try to brute force and just perform experiments . . . just to find out rather than trying to do the computation (AI17).

Five interviewees discussed use of AI/ML to identify patterns or phenomena which human beings could not in data. This utility is valuable across feature recognition, measurement of previously known constructs, and modeling of unknown relationships. AI/ML can also be used to measure known constructs based upon complex and multidimensional data. AI19 provides an example of measurement:

> [W]e send a short pulse of electromagnetic radiation onto some object, some of it . . . bounces off, some of it goes through and bounces off subsequent layers in the structure. So if the layers are close together, the kind of echoes . . . overlap might be so severe, two echoes from successive layers might actually look kind of like one echo, and there will be subtle differences between an echo from a single layer and overlapping layers, echoes from two layers, and it's hard to know what those distinguishing features are beforehand. So AI techniques, in essence, can automatically find those specific signatures that will kind of differentiate between a double echo that's highly overlapping and a single echo, and will be able to give us the thickness of that layer (AI19).

Multiple interviewees discussed the value of using AI/ML to model unknown relationships:

> [The] majority of the, let's say, physics models in terms of mathematical format, they are summarized by people, right? But the thing is, if the patterns are not seen by people, it's difficult for them to summarize into the mathematical format, right? So since we are not able to see easily by people [sic], we want to use machine learning, because machine learning is very efficient to see such pattern [sic], especially a mass amount of data, right? (AI05).

> We look at phase diagrams, and there's a lot of information in a phase diagram, but converting that into useful material property predictions isn't humanly tractable. Because . . .you have a qualitative image, you're asking for a quantitative value. And so there's obviously a ton of variance that you get if you ask somebody to do that by hand. And so what we did is, we trained a neural network . . . to predict that quantitative value from a qualitative image, right? And that's something that it can do, and we showed that it worked very well . . . for the broad range of materials that we look at (AI33).

However, one interviewee was somewhat skeptical of this utility, at least in certain domains:

> I always tell my students . . . if you can't solve it with a simple machine learning thing like a linear equation, if you can't see a slight benefit or something going on there when you solve it simply, then it's probably not going to be solved with something complex. . . Like, you might try to solve the problem simply, and then you can't see the relationship, but then it forces you to take a step back and say, well, maybe the problem is I'm looking at too many different situations here. Like I'm looking at this machine, as well as when it's off, as well as when there's, you know, all these three operators . . . Maybe I should focus on just when operator A is working, and when the machine is working, and, you know, when the temperature is above 70 degrees or something. Yeah. And now I'm narrowing my focus. Now is there something there? So rather than . . . throwing math at it, you know, try to understand the situation first. . . And there are those rare cases where I'm stuck, but I'm going to throw all the math at it, and yeah, something pops out. But those cases are very rare (AI08).

Finally, four interviewees stated that AI/ML use can improve the quality of research, all in terms of accuracy and reliability:

> We're still sort of working at the same cost as the standard quantum mechanical technique, we're still doing quantum mechanical simulations. We're just using machine learning to improve some of the approximations (AI11).

To summarize, then, participants most frequently discussed benefits of AI/ML in research in terms of reducing accuracy/cost or accuracy/speed trade-offs in research, and particularly in computational modeling. Several interviewees also stated that AI/ML permits solution of novel problems, and a few stated that AI/ML can, when used as a supplement to preexisting modeling techniques, improve modeling accuracy.

*4.4 Downsides of artificial intelligence or machine learning use*

Interviewees discussed a variety of downsides or weaknesses to AI/ML use in research. Most commonly discussed were problems of accuracy, addressed by 24 interviewees. While AI/ML tools can help to decrease computation cost/accuracy trade-offs, they are still often not as accurate or reliable as conventional, physics-based models. Interviewees noted variously that, in addition to providing inaccurate predictions, AI/ML models can do so unpredictably.

Interviewees were adamant that AI/ML-based tools, being high-parameter models fit based upon data, depend heavily upon their training data. Interviewees stated that they have limited ability to extrapolate beyond the bounds of the datasets on which they were trained, sometimes in contrast to conventional theory:

> [O]nce you establish that fundamental relationship . . . it extrapolates well. You use *F=ma* for a golf ball, you use it for a planet, it largely works right? . . . [D]ata driven models so far don't have that same ability (AI13).

> [Y]ou're only able to interpolate . . . This is based on the type of equations and, and packages that we're dealing with, which are extremely nonlinear . . . [I]t will give you wrong results, if you try to use the present data, and approaches to extrapolate things beyond the range of the data you have (AI14).

Several interviewees also noted the importance of high-quality data for effective use of AI/ML, repeating the adage, "garbage in, garbage out." Accordingly, six interviewees emphasized the importance of expert supervision and checking of AI/ML systems in research; models' results cannot be taken on faith.

Beyond accuracy issues, 11 interviewees stated that AI/ML models can be difficult to interpret. When we probed what this meant and why it mattered, interviewees observed that it can be difficult to predict AI/ML models' behavior and performance under different circumstances, thus, to judge their situational utility or to repair errors.

> Here's the problem. If I change it, if I have a slightly different machine or if I'm using a slightly different material . . . is it still going to work as well? And the answer is, I don't know. . . If you have a physics based system, we say, yeah, I can change all these material characteristics and it will work well . . . Or you can say no, because I know it's these particular characteristics that allow me to use this approach . . . But I can give you a definitive answer (AI27).

> I've seen cases where it's going fine for a very long time, 500,000 steps, and suddenly . . . The molecular dynamics randomly comes to a configuration that doesn't really match the training data set, and an atom suddenly goes from moving at, like, meters a second to kilometers per second, and your entire simulation just completely failed. And you don't necessarily know exactly why that happened (AI34).

Interviewees observed that these interpretation difficulties can make it difficult even to assess the accuracy or limits of AI/ML models, or to establish their superiority or the validity of their results compared to other research methods.

In a final "core scientific" concern, five interviewees offered the opinion that AI/ML lacks the ability to develop novel scientific theory:

> In terms of the papers that are really making an impact out there and some of the good fundamental work, you know, that are advancing our knowledge base and not just about which algorithm is the best. . . I mean, when you get to the higher end research, you're really extrapolating. . . It's just right out of the realm of AI. AI is all about, you know, interpolating and operating within known, trained data set (AI27).

This is not inherently a problem – after all, human researchers can still develop novel theory. However, some interviewees were concerned that, by allowing researchers to "skip" processes of theory development or extensive courses of experimentation, AI/ML may hinder discovery in the long run. One researcher specifically discussed missing out on productive anomalies which might be discovered through extended experimental campaigns:

> Let's say that you want to develop this alloy. . . I can probably just do three samples, five samples rather than 100,000 samples . . . and let the machine learning tell me what the final alloy is. And then you're missing a lot of different alloys or maybe optimal remedies that could have existed, that could have found, if you did more experiments. . . I don't know if it's going to help or it's going to impede progress in science, in the long term. . . (AI14).

Others noted that superior alternatives or fruitful anomalies might be found even with high-fidelity physical modeling but missed with AI/ML. In short, while many interviewees suggested AI/ML can reduce the resource intensity of research to solve specific, near-term design problems, some voiced concerns about their impact on scientific exploration and theory development.

Although AI/ML tools bear potential to decrease the resource intensity of at least some research, several interviewees noted that the reverse can occur. Though AI/ML tools are often computationally cheaper than equivalent physics-based models to run, they require substantial computation time to train in the first place. In addition, acquiring and cleaning the vast quantities of data required for this training is often an expensive and labor-intensive process:

> We've been able to bring times and costs of the amount of experiments or simulations that we need to do to get a good model . . . 30 to 50 percent kind of reductions in the amount of effort. But right now that 30 to 50 percent gets eaten up on the front end of actually designing, how are we going to generate the data? How are we going to analyze the data in

> this real, pseudo real-time fashion? There aren't a lot of general methods for doing that yet, and that's where a lot of the innovation needs to happen to really, truly realize. You know, okay, we had to do 50% less experiments, but we had to do 100% more work up front right now, right? We'd like to get to the point where you really, no more effort up front and you really start to realize those benefits (AI09).

Two interviewees were also vocal about the energy intensity of training (and sometimes of running) AI/ML models, and the environmental consequences thereof

> These AI methods are so energy inefficient. . . I'm torn because as somebody who's in this space, I'm responsible to understand how they work. . . At the same time . . . I am hit with some guilt at . . . how much energy I'm consuming and how much, what that translates to in terms of fossil fuels and carbon footprint" (AI32).

Thirteen interviewees observed that AI/ML tools are very easy to use inappropriately. Several interviewees opined that collaborators and colleagues often are not familiar with the specific affordances, requirements, and weaknesses of AI/ML models, and that this can lead to wasted time and resources; or simply poor research:

> [T]here's always this fear I have that people are too trusting in ML and AI type models, versus, you know, somebody tries to say, well, I want to understand heat transfer of a cow. Well, so let me start by assuming a spherical cow. Well, you know, that there's kind of a big assumption up front, right? So if things don't work too well, then you say, well, maybe we need to revisit that initial assumption. You know, I think there's a danger that people don't do that with machine learning. . . . If all you have are like this simple polynomial model and data, you sometimes forget that, gee, if the conditions change too much, then my stuff won't work. And I think you're more tuned into that if you're building models based upon some more science-based approach (AI31).

However, this problem was not necessarily unique to AI//ML systems:

> It's very common in our field for me to even read a paper where someone has used a physical simulation and go, I don't trust that they didn't think carefully enough about what their error is and what their conclusions are and how their conclusions depend on the error. So I think that that pattern has just continued with ML. People want a number and they'll get the number from physics or they'll get it from an ML. And if it confirms what they want to show in the paper, then they call it good (AI11).

Several interviewees noted the faddish nature of research tool trends, suggesting AI/ML may be used when not appropriate or necessary:

> I've been in the field long enough to see trends, where people will often use things because they're fashionable, but not necessarily because they make sense. For example, about 15 years or so ago, there was a huge trend in using genetic algorithms to do all types of optimization. And it seemed like people just thought if they used a genetic algorithm on their problem that they would get the paper published. And genetic algorithms are only appropriate for a very small subset of problems that have certain characteristics. Otherwise, there are other algorithms that you really should use that'll be much faster and much more computationally efficient. . . I think we're seeing similar things right now in artificial intelligence. . . In many cases, [researchers] may be using machine learning algorithms like neural networks, for example, to learn relationships between variables when there are other techniques that have been around for a long time that might do a better job or with less data (AI26).

Several interviewees suggested that AI/ML techniques should primarily be used for problems which could not be solved using previous methods; and that longer-standing methods are often superior when available.

## V. Discussion: Another Rung on the Ladder of Testing Methods?

Computation has longstanding applications in MMS research, and indeed in engineering more generally. It has long been used for "vicarious trial," i.e., for assessing the performance of possible system designs outside the context of real-world practice. Computational statistical methods, even as simple as linear regression, have also been used to help researchers to develop theory to describe and predict the performance of phenomena of interest. Finally, mathematical optimization methods have helped to "suggest" high-performing system designs or configurations.

Among our interviewees, for purposes of computational modeling, AI/ML does not expand beyond these existing utilities. However, it extends these capabilities substantially. First and most obviously, "surrogate modeling" reduces the computational cost of vicarious trial for phenomena already described by reliable, mathematically expressed theory, permitting researchers to search larger design spaces more rapidly or even to model systems or phenomena which were previously too computationally intensive to model. Second, AI/ML systems can be used to model and predict phenomena not robustly covered by existing theory, either as a corrective to remaining uncertainty within an existing model or as the entire model. In such cases, the researcher surrenders the actual task of theory formulation to the model. The “theory” developed as an assemblage of model parameters has not been explicitly designed by humans and may not be practically applicable or even fully interpretable by humans.

**Figure 3:** Relationship between ML-based modeling, conventional physics-based modeling, and physical trial.

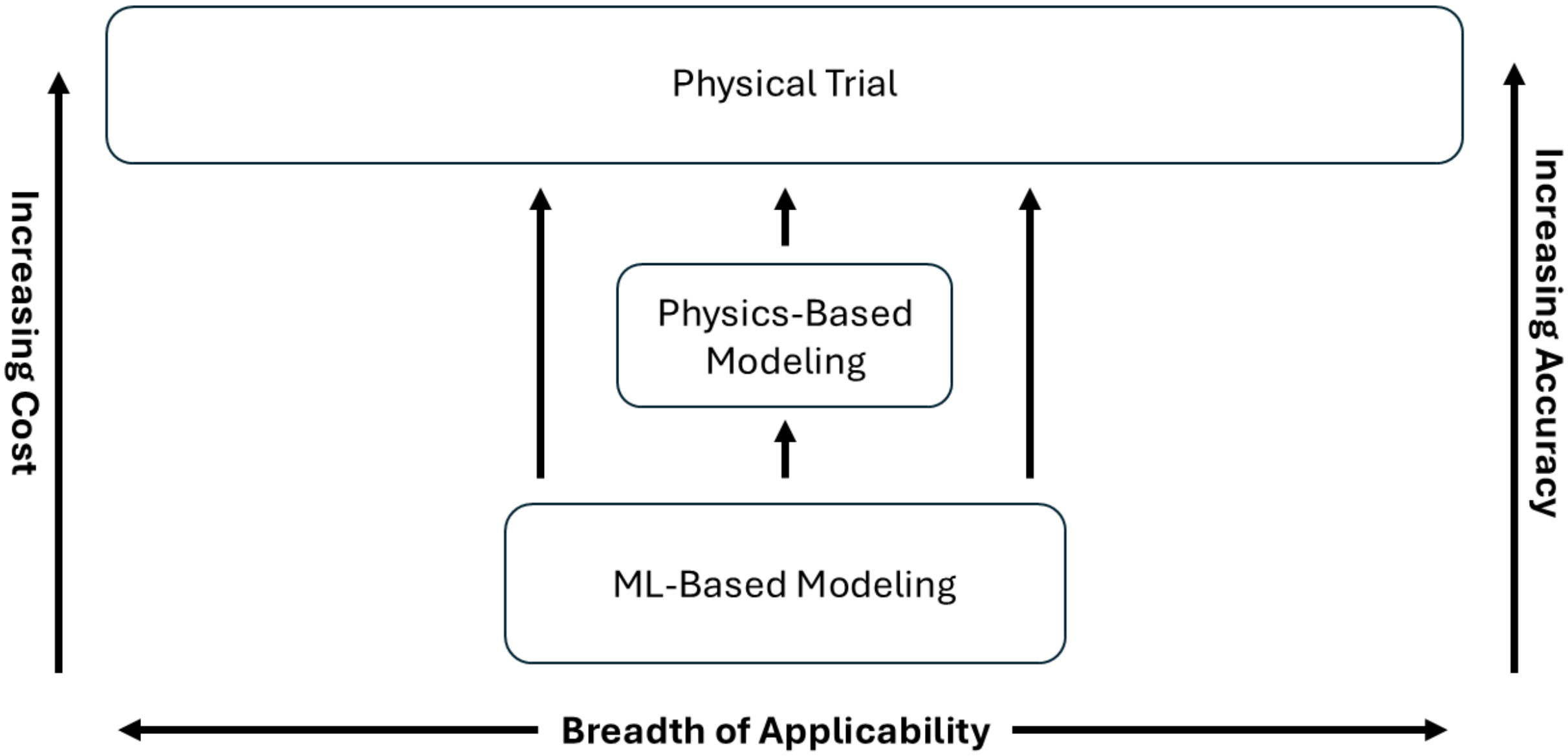

Vicarious trial has inescapable limits in reliability. The essence of theory is abstraction – a simplified representation of real-world phenomena (Messeri & Crockett, 2024). Theory achieves its power by leaving out many details deemed, hoped, or overlooked as irrelevant to the outcomes of interest. For high-stakes applications, modeling can never replace practical trial; its function is in allowing researchers and engineers to (more) rapidly sift through plausible alternatives, to find those most promising to explore through live trial. It is very valuable, as Yaqub and Nightingale (2012) observe, to have a “ladder” of models, ranging from the abstract, simple, and easy to apply, and stepping through can be made cheaper. It also is applicable over a greater breadth of phenomena than conventional modeling, as it can be applied to phenomena for which physical theory currently does not exist.

increasing levels of complexity and resemblance to real-world practice.[2] Surrogate modeling via AI/ML adds a cheaper “step” below the reliability and computational intensity of some existing physics-based modeling approaches, such as density functional theory. While less reliable, surrogate modeling permits rapid exploration of design spaces to identify candidates worth passing on to physics-based modeling (and hopefully some subset thence to practical trial).

Meanwhile, for areas lacking existing theory, AI/ML techniques can add a new “step” of simplified, computational modeling below physical trial, which was heretofore the only option. We depict the relationship between ML-based modeling, conventional physics-based modeling, and physical trials in Figure 3. ML-based modeling can either directly model physical phenomena, identifying promising candidates for physical trial; or surrogate model physics-based models, identifying promising candidates for conventional physics-based modeling. ML-based modeling is typically less accurate than physics-based modeling

When used for surrogate modeling, ML-based modeling is inherently cheaper and less accurate than the physics-based models for which it proxies. However, when used directly to model poorly-understood physical phenomena for which physical theory is not available, certain novel properties appear relative to conventional, computational modeling. Because the "theory" encoded in ML models of poorly-understood phenomena was developed via ML rather than explicitly designed by a human, practitioners and developers may lack understanding of when and why the ML-based model is or is not accurate and reliable. These downsides constitute "epistemic risks" – or risks of error, ignorance, or other kinds of failure in knowledge-production activities (e.g., Biddle & Kukla, 2017; Biddle, 2022; Messeri & Crockett, 2024).

[2] This is not to speak only of computational modeling. Computational modeling techniques typically sit somewhat lower down the “ladder” than do scale models or controlled experiment techniques. However, there can be many layers of practical trial between even the laboratory and real-world practice. For example, even once a novel chemical has been synthesized and found useful, a large-scale “test plant” is often required to work out the chemical and logistical problems of producing it at industrially useful scale (Rosenberg, 2000).

Many interviewees stated that ML-based models are reliable only in regions of dense training data – a helpful heuristic. However, this limit reduces ML-based models' exploratory potential and reliability as compared to more conventional theory. It is certainly possible that ML-based models can, via pattern recognition, internalize poorly-understood physical phenomena and provide novel extrapolation capabilities. However, they will often produce incorrect extrapolations. It cannot be known whether they have found something real save via checking with a researcher's hard-won "physical intuition," more robust theory-based models or, ultimately, via practical trial.

Beyond modeling, AI/ML techniques are also lowering vicarious trial costs in other ways. Pattern-recognition techniques permit automated preprocessing of data, such as visual data or the results of chemical composition analyses, which previously would have required labor-intensive and sometimes cognition-intensive manual analysis. Furthermore, though it was little discussed in our sample, it is possible to use AI/ML in combination with robotics to automate the manual operations involved in certain research processes. Such automation of data preprocessing and of experimental tasks themselves can help to decrease the time, materials, and labor costs of physical trial itself.

In sum, we find that a qualified "yes" in answer both to RQ1: "*Is AI improving the breadth, reliability, or cost of vicarious trial?*" and to the subset RQ2: "*Is AI taking over research tasks previously performed by humans?*" AI/ML permits modeling of previously unmodelable phenomena, decreases the cost-reliability trade-off in modeling, and can be used to directly increase the reliability of existing physics-based models. In part, it can do this by taking over the previously human task of designing and tuning computer models of phenomena, allowing computer models of phenomena not currently understood or modelable through human theory to be constructed. In addition, AI/ML can automate and cheapen components of the physical experimental process, such as data preprocessing and preliminary interpretation. That is, AI/ML can improve the breadth, reliability, and cost of vicarious trial by cheapening and extending the utility of computational modeling; and by automating, cheapening, and speeding certain experimental tasks.

This is not, however, to say that AI/ML has unlimited utility. Many interviewees indicated that AI/ML-based models are most useful for interpolation of phenomena within well-mapped physical domains, and that they have not (as yet) demonstrated the capability to develop novel, powerful theory capable of major predictive extrapolations – the sort of novel, parsimonious, and general theory by which qualitatively new scientific and technological opportunities may be discerned. Several of our interviewees expressed severe doubts about this possibility. Moreover, they expressed concerns that use of AI/ML for short-term gains would rob researchers of opportunities to make disruptive advances themselves. In short, mechanistically, AI seems best suited to "sustaining innovation" within existing technological paradigms. In entrenching existing paradigms and focusing effort within them, AI may have limited or even negative effects on society's ability to produce disruptive new technologies (Dosi, 1982; Dosi, 2007; Nelson, 2008b).

Our finding that AI/ML is indeed being used to cheapen and speed vicarious trial aligns well with the limited prior literature on contributions of AI to scientific and technological development. From general discussion, our findings on reported specific applications and utilities of AI/ML in science align with certain of Krenn and colleagues' (2022) collected anecdotes on AI for revealing phenomena and data representation. Our findings on the importance of resource efficiency align with Besiroglu and colleagues' (2024) finding of increased research productivity for computation Meanwhile, our finding that AI/ML suggestions ultimately must still be validated by practical trial aligns with Agrawal and colleagues' (2024) simulation-based argument that AI/ML gains to technological progress will depend heavily upon capacity to test its suggestions.

Our interviewees' concerns about "losing touch with the phenomenon" and taking shortcuts around potentially fruitful but laborious scientific work align with certain of Messeri and Crockett's (2024) concerns about epistemic degradation in AI/ML-enabled science. Depending upon theoretical framing, our findings may be taken as consonant with Bianchini and colleagues' (2026) bibliometric finding that AI adoption generally associates positively with impact (measured by follow-on citations). However, they may be in tension with Bianchini and colleagues' (2026) finding that AI adoption generally associates with scientific article novelty (measured by new word combinations)[3], and

[3] We harbor concerns that the word-cooccurrence-based measure of novelty used in this impressive econometric study may be only a weak proxy for scientific advancement, particularly given the measure's positive correlation with

their argument that both occur thanks to improvements in the process of hypothesis formulation and recombinatorial search in complex knowledge spaces. However, as noted previously, there are vast and perilous gulfs between linguistic novelty, scientific novelty, and technological advancement; and the linguistic space of a scientific field's papers is not the same thing as the landscape of technology and practice (Bonvillian, 2018; Stokes, 1997). In short, our findings provide novel evidence on the nature of, reasons for, scope of, limitations on advantages offered by AI/ML in technology development, extending prior theory on processes of technological change and informing conditional expectations of AI/ML's capabilities.

In the context of the broader innovation studies literature, our findings support and extend existing theory on the role of scientific research in technological change. It has long been recognized that science is, in Toulmin's (1966) phrase, an overhead on technological development, providing proxy methods of experimentation which permit cheaper search through technological possibility landscapes; and suggesting directions in which to search. As primarily a set of new analytical techniques, AI/ML does not break this pattern. Rather, its utility can be summed up: more "maps," and more mapmaking tools. Nor does AI/ML directly provide new observations (and, despite the possibility of self-driving labs, particularly not new *types* of observations).

Rather, AI/ML helps to more intensively exploit those data which have already been and are already being collected—including, crucially for our interviewees, disparate experimental runs and the vast and largely unexploited data streams pouring from modern, computerized manufacturing processes (to say nothing of the rest of our increasingly sensor-suffused world). Cheaper and faster modes of analysis, through surrogate modeling, allow technology developers to filter through possibilities more rapidly, which in turn allows them to increase the number of possibilities considered. Phenomenological modeling and supplementation of physics-based models may permit scientific foresight to reach further from the foundation of experience than has previously been achieved, extending guidance a little more into the endless frontier of uncertainty—but, so to speak, toward inward frontiers rather than outward.[4]

Because it depends upon historical circumstances and heavily benefits from built-up material and intellectual infrastructure, the process of technological change is highly path-dependent. This path dependency can be summarized in Dosi's (1982, 2007) concept of "technological paradigms," i.e., built-up enterprises of material artifacts, vocabulary, and habits of thought and action which focus and facilitate innovative effort consonant with those preexisting enterprises. Most innovation is "sustaining," improving and entrenching existing technological systems; paradigms are reformulated and redirected by occasional bursts of "disruptive innovation" (Christensen et al., 2018). Systematic and steady experimentation, development, and improvement are not feasible, nor even definable, without at least a loose paradigm (Nelson, 2023a; Nelson, 2008b). But returns to innovative effort within any given paradigm often diminish over time (Dosi, 1982; Dosi, 2007; Nelson, 2008b; Perez, 2009). Entrenchment of dense and highly improved current paradigms can make disruptive invention and implementation less likely, more difficult, and more costly (Cowan, 1990; Hughes, 1987; Trencher et al., 2020).

Our interviewees' caution that AI/ML is very powerful for interpolation and relatively poor at extrapolation suggests its utility is, for the time being, limited to helping us to predict and control phenomena which are already extensively recorded, even if poorly understood. In fairness, there is already at least one exception. As have prior statistical techniques, AI/ML can help to plan campaigns of experimentation—in the idiom of science-as-map for technological search, in suggesting which future voyages may contribute effectively to the cartographic enterprise. Yet maps of mapmaking, too, must be incomplete and imperfect. If the total number, coverage, or diversity of experimental campaigns decreases, so too do opportunities to discover qualitatively novel opportunities or anomalies. To risk undue speculation, there is a promise and a danger that AI/ML may be a sustaining, thus entrenching, innovation to the process of sustaining innovation itself, circumventing swaths of laborious empirical exploration in which investigators might stumble over unforeseen and disruptive opportunities (Dosi, 1982; Nelson, 2008b). Yet whether or not this suggestion is borne out, AI/ML will, like previous analytical techniques, fields' preexisting states of linguistic complexity. In plain language, we are concerned that fields containing high linguistic diversity may be more prone to produce novel word combinations in general, and that this may indicate, rather than advancement, merely casting about under low paradigmatic consensus. Our concern is intensified by the facility of large language models in assembling combinatorially novel (at least for a given scientific subfield) turns of phrase. But assessing the validity of these concerns will require further research on the nature and measurement of scientific novelty.

[4] There are likely to be differences in the long-run usability and fruitfulness of "theory" constituted by explicit, parsimonious, and human-interpretable language and mathematical expressions versus vast arrays of parameters stored on hard drives, but detailed exploration of this possibility must be left to future inquiry.

continue to depend upon robust foundations in and feedbacks with real-world experimentation and observation.

## VI. Limitations and Conclusions

Though AI and ML have attracted mass attention only recently, MMS researchers have for decades been steadily, incrementally developing and applying statistical and computational tools, in particular ML. Recent advancements in computational capacity and techniques have permitted these tools to take on ever-wider roles in supplementing conventional theory-based models of MMS phenomena – or, when theory is not available, substituting (in part) for it. Our findings suggest that, rather than some qualitatively novel method of scientific discovery or invention, contemporary AI/ML tools in science expand upon the utility of prior forms of computational modeling. They are best thought of as adding another "rung" to the ladder of vicarious trial in some areas. They provide another method of predicting the viability of different technological alternatives, at lower cost and with lesser accuracy than existing, robust, physics-based models; and a way to predict, at unclear accuracy, phenomena in domains where robust physics-based models are not yet available. Situated within prior evolutionary theory on technological change, this utility is important and valuable, but simultaneously contingent and limited. AI/ML's ability to permit faster and cheaper development, evaluation, and spread of technologies depends upon its situation within a robust preexisting system of empirical data collection, conventional analysis, and theory development.

A major outstanding question, which likely cannot yet be answered, is whether AI/ML-based tools are capable of disruptive novelty. A hallmark of previous major, technology-relevant advancements in science – Newtonian mechanics, germ theory, the theory of the atom, the structure of DNA – is that they have facilitated prediction and control of very wide ranges of phenomena, some of which were outright unsuspected prior to these theories' development. Within certain domains and with certain priorities, these theories "travel well" – they permit parsimonious summary of complex phenomena, and prediction of technological opportunities of new varieties. Sterilization, systematic vaccine development, pasteurization – these interventions were nearly unintelligible without germ theory. Are AI/ML tools capable of such novelties? Interviewees who commented on the question suggested that they are not. Rather, they asserted that AI/ML models travel rather poorly across phenomena, and particularly poorly into data-sparse domains.

Indeed, AI/ML "explanations" of recalcitrant phenomena are human-uninterpretable arrays of values within neural networks, almost inevitably overfit to specificities of their training datasets and conveying no generalizable functional form. They may point the way to unrecognized phenomena – a valuable utility – but it is not yet clear that they can facilitate parsimonious prediction or control of such novel phenomena. Nor is it clear that the implicit "theory" developed within AI/ML models can enter into the recombinant process of extrapolation and inspiration that has produced such advances as Darwin's reapplication of Malthus to develop the theory of evolution, or the adoption of physics and chemistry methods into biology as biophysics and biochemistry. Can, say, an ML "theory," e.g., an overfit array of parameters developed on a titanium alloy dataset, suggest anything novel about biology, sociology, or economics—or, indeed, vice versa? The prospect does not seem likely at present.[5] As discussed above, it is not yet clear even whether particular ML models can translate well from one material to another. And, of course, theory and analysis are only two, valuable but insufficient, tools in the vast toolbox required for technological advancement. Mighty observational, organizational, infrastructural, and translational capacities must be maintained and mobilized if words and figures are to durably improve widespread artifacts and practices.

These results suggest that research strategists and policymakers should be cautious about claims of AI/ML fundamentally revolutionizing science and technological development. For our interviewees, AI/ML are just more tools in the toolbox – powerful in certain applications, but with their own weaknesses and limitations. Our interviewees suggest that AI/ML are most useful in ensemble with the other, often more reliable methods of "vicarious trial" which predate them, and which step along the ladder from simplified testbed to real-world practice. Our interviewees' doubts about the power of AI/ML to discover or usefully characterize qualitatively novel, general phenomena provide further reasons for caution. So does their concern that AI/ML could provide a "shortcut" around slower experimentation and theory development, which may be more fruitful in the long run. In short, AI/ML can certainly benefit research and development. But they require skilled users, and integration within robust systems of more conventional theory development and technological trial. We will continue to need expert technologists, explicit, human-interpretable theory, and empirical trial for the foreseeable future.

These findings, in both their promise and their limits, carry important implications for research policy and strategy. By our account, AI/ML as analytical method can contribute fruitfully to science as one more element of the superstructure built to exploit the results of experiment, data collection, and prior theory. AI/ML can permit researchers to make more of the data they have, but it does not wholly substitute for any preexisting element of the scientific enterprise. These data analysis techniques depend crucially upon well-functioning and fruitful experimental, data collection, conventional theoretical, and conventional computational methods. In MMS, a diverse and flourishing research ecosystem including all these subcomponents will be required to maintain technological progress, and, indeed, to maintain the utility of AI/ML itself.

[5] Meanwhile, large language models can certainly draw analogies between disparate areas of science. But only a human expert and, ultimately, empirical trial can say whether such speculations are fruitful.

Accordingly, research funders and policymakers should regard investment in AI/ML infrastructure, human resources, and scholarship as a complement to and multiplier for, rather than an independent substitute or replacement for, conventional elements of the research enterprise. Research managers and sponsors should ensure attention to whole workflow operation and cross-disciplinary collaboration when funding and supporting the adoption of AI/ML technologies and provide greater incentives for high-risk research to ensure that researchers are encouraged to pursue disruptive innovation. To adopt Messeri and Crockett (2024)'s agricultural metaphor, a robust, resilient, sustainable, and productive research ecosystem depends upon diverse, interwoven, and mutually fertilizing branches of research, in terms of topics, data collection methods, analytical techniques, and theoretical approaches (Alic et al., 1992; Fuchs, 2021; Nelson, 1961). Investment in AI/ML is merited, but monoculture would lead to sterility.

More, of course, remains to be learned about AI and ML in research. While our interviewees often boasted decades of experience at the cutting edge of MMS research, they nonetheless constitute a modest sample, situated at major U.S. universities, within only one general field of research. In addition, the modest response rate makes selection effects plausible, though the specific nature of such effects is unclear. There are reasons both to expect higher enthusiasm and more critical interest in the high-level effects of AI/ML in research among respondents, but this is mere speculation. Our results cannot speak directly to developments in industry, in government labs, other nations, or, perhaps most importantly, to other fields of research and development. Exploratory analysis showed no strong association between MMS subfields and particular benefits or downsides, with all observations well represented across subfields (Appendix 3). This consistency shows promise for cross-field generalization of our findings. Nonetheless, particularly given the modest-per-subfield sample size, this is yet promise only.

Further interview studies over time and expanding along each of these dimensions will help to clarify the scope and generalizability of our findings. Several of our interviewees suggest that AI/ML tools in R&D may not be developing so rapidly as has sometimes been suggested. Nevertheless, they are developing quite rapidly; we can at best provide only a snapshot of a dynamic field. Beyond interview studies, there is a pressing need for additional empirical study of this topic using all methods, from bibliometric analyses of publication and invention rates and contents; to ethnographic or interview study of the on-the-ground usage and detailed consequences of AI/ML and automation in research. Survey studies of practicing researchers would serve similar goals, and permit examination of factors affecting different views on and experiences with AI/ML.

We hope our study may also illustrate the value of contextualizing study of AI/ML within extant theories of technological change. A detailed understanding of knowledge creation and technological development clarifies what AI/ML will need to do to genuinely transform these processes; and what its consequences may be. It is all too easy to abstract from scientific and engineering processes to view technological progress as a commodity product to be stockpiled. Yet technological change is fundamentally a process of *replacement*--of altering communities', organizations', and societies' methods of organizing themselves and arranging matter to pursue their goals. Moreover, each increment of technological progress is something different from those before – the greater the increment, the greater the difference. Understanding technological change as a *search* process, or a *variation-and-differential-selection* process, redirects focus from the quantitative, but intermediate, products of science and technology – most prominently, publications and patents – to whether and how AI and ML are altering how technological communities learn.

## APPENDIX 1: SEMI-STRUCTURED INTERVIEW PROTOCOL ON ARTIFICIAL INTELLIGENCE IN RESEARCH

1. *Personal background*
   1. What is your current position? [Only if unknown]
   2. How long have you worked in engineering research or development, and how long do you expect to continue?
   3. Please briefly describe your general area of research.
2. *Personal use of AI*
   4. Do you consider yourself to work with AI? If so, how long have you done so?
   5. For what have you used AI in your research and development, and why?
      a. Have you experienced any benefits to use of AI in your research? If so, what?
      b. Have you experienced any downsides to use of AI in your research? If so, what?
   6. Has AI affected what topics you choose to research? If so, how?
   7. Has AI affected funding, career, or collaboration opportunities available to you as a researcher?
   8. Has AI affected what skills you need to conduct research? If so, how?
      a. Do you think AI has eliminated skills you would have needed to perform research otherwise?
   9. Has AI affected what human or material resources you need to conduct research? If so, how?
   10. Has AI affected how much or how rapidly you can conduct or publish research?
   11. Has AI affected the quality or reliability of your research?
   12. Has AI affected how you communicate or access research?
      a. For example, has it altered the form of your research outputs?
   13. If another researcher uses AI in a study, does this affect whether you can rely upon or replicate the study's results? If so, how?
   14. Has AI affected who can use or benefit from your research?
   15. Would you like to share any other thoughts on your experiences with AI in research?
3. *Wrap-up*
   16. Whom else should I interview on this topic?
   17. Do you have any other questions or comments for me?

## APPENDIX 2: RESPONDENT INFORMATION

**Table 5:** Respondents' university affiliations. Italicized universities are not among the top 10 most productive of relevant Web of Science papers, discussed above. These respondents were identified at manufacturing and materials science conferences.

| **University** | **Respondents (n=32)** |
|---|---|
| Georgia Institute of Technology (Georgia Tech) | 14 |
| Purdue University | 3 |
| University of Michigan | 3 |
| Northwestern University | 2 |
| Virginia Tech | 2 |
| Rutgers New Brunswick | 1 |
| Texas A&M University | 1 |
| *Auburn University* | 1 |
| *Harvard University* | 1 |
| *University of Chicago* | 1 |
| *University of Illinois, Chicago* | 1 |
| *University of Tennessee, Knoxville* | 1 |
| *University of Wisconsin, Madison* | 1 |

**Table 6:** Respondents' ranks.

| **Rank** | **Respondents (n=32)** |
|---|---|
| Full Professor | 12 |
| Associate Professor | 7 |
| Assistant Professor | 2 |
| Research Faculty | 4 |
| Postdoctoral Fellow | 4 |
| PhD Student/Candidate | 2 |
| Recent PhD Graduate | 1 |

**Table 7:** Respondents' genders. Gender categories with zero respondents are omitted.

| **Gender** | **Respondents (n=32)** |
|---|---|
| Female | 4 |
| Male | 28 |

**Table 8:** Respondents' fields of PhD study. Very similar field names (e.g., "Materials Science" and "Materials Science and Engineering," or "Industrial Engineering," "Industrial and Systems Engineering," and "Industrial and Operations Engineering") are coded together.

| **Field** | **Respondents (n=32)*** |
|---|---|
| Mechanical Engineering | 13 |
| Materials Science and Engineering | 9 |
| Industrial, Systems, and Operations Engineering | 5 |
| Electrical Engineering and Computer Science** | 2 |
| Chemical Engineering | 1 |
| Physics | 1 |
| Scientific Computing | 1 |
| Theoretical and Applied Mechanics | 1 |

*One respondent held a dual PhD, so, while the number of respondents is 32, the number of fields is 33.
**This is not a consolidation; two respondents explicitly held PhDs in "Electrical Engineering and Computer Science."

## APPENDIX III: THEMATIC RESULTS BY SUBFIELD

**Table 9:** Modes of artificial intelligence and machine learning use by subfield. All codes nonexclusive (n=32).

| | **Manufacturing process design, characterization, and control (n=18)** | **Materials design or characterization (n=15)** | **Product design (n=5)** |
|---|---|---|---|
| AI/ML modeling of phenomena | 14 (78%) | 10 (67%) | 4 (80%) |
| Phenomenological modeling of poorly understood phenomena without explicit theory* | 14 (78%) | 7 (47%) | 3 (60%) |
| Surrogate modeling of physics-based models | 7 (39%) | 6 (40%) | 1 (20%) |
| Corrections for physics-based models | 3 (17%) | 0 | 0 |
| Planning experimental campaigns | 1 (5.6%) | 3 (20%) | 1 (20%) |
| Measurement | 2 (11%) | 1 (6.7%) | 0 |
| ML design tool as research product | 0 | 0 | 1 (20%) |
| Self-driving lab | 0 | 1 (6.7%) | 0 |
| Large language models** | 6 (33%) | 8 (53%) | 3 (60%) |
| General literature search and review | 3 (17%) | 6 (40%) | 2 (40%) |
| Drafting papers or preparing graphics | 3 (17%) | 3 (20%) | 3 (60%) |
| Extracting data from documents*** | 0 | 4 (27%) | 0 |
| Assistance with programming | 0 | 1 (7%) | 1 (20%) |
| Translating papers | 0 | 1 (7%) | 0 |

*A large share of interviewees' statements about modeling which did not explicitly pertain to surrogate modeling likely refer to phenomenological modeling. However, to avoid overinterpretation, we have assigned this code only where the interviewee explicitly discussed using AI/ML to model unknown relationships.
**The interview protocol did not focus on large language models, so LLM use is likely underreported in our sample.
***This application is the strongest case for an across-field difference, as it occurred only in materials design and characterization. However, given the small field-specific sample sizes and the fact that this study does not focus on LLMs, we pass over this result.

**Table 10:** Benefits and downsides of artificial intelligence and machine learning in research by subfield. All codes nonexclusive (n=32).

| | **Manufacturing process design, characterization, and control (n=18)** | **Materials design or characterization (n=15)** | **Product design (n=5)** |
|---|---|---|---|
| ***AI or ML benefits*** | ***16 (89%)*** | ***13 (87%)*** | ***5 (100%)*** |
| Reduction in accuracy/cost/speed trade-off in research, especially computer modeling | 14 (78%) | 13 (87%) | 4 (80%) |
| Address of previously unsolvable problems | 9 (50%) | 6 (40%) | 4 (80%) |
| ***AI or ML downsides*** | ***18 (100%)*** | ***15 (100%)*** | ***5 (100%)*** |
| Accuracy weaknesses | 13 (72%) | 13 (87%) | 3 (60%) |
| Interpretability weaknesses | 8 (44%) | 5 (33%) | 2 (40%) |
| Long-run scientific progress concerns | 5 (28%) | 3 (20%) | 1 (20%) |
| Resource issues | 5 (28%) | 3 (20%) | 1 (20%) |
| Inappropriate use issues | 7 (39%) | 9 (60%) | 1 (20%) |